\documentclass{PoS}

\def\mg{MG}
\def\rmsub#1#2{#1_{\mbox{\tiny #2}}} 
\def\secspace{\vspace{-0.38mm}}

\title{Adaptive Multigrid Algorithm for the QCD Dirac-Wilson Operator}

\ShortTitle{Adaptive Multigrid Algorithm for the QCD Dirac-Wilson Operator}

\author{J. Brannick$^a$, R. C. Brower$^{bc}$,
  \speaker{M. A. Clark}$^b$, J. C. Osborn$^{bd}$ and C. Rebbi$^{bc}$\\
  \llap{$^a$} Department of Mathematics, The Pennsylvania State
  University, 230 McAllister Building,
  University Park, PA 16802, USA\\
  \llap{$^b$} Center for Computational Science, Boston University, 3
  Cummington St,
  MA 02215, USA\\
  \llap{$^c$} Department of Physics, Boston University,
  590 Commonwealth Avenue, Boston,  MA 02215,  USA\\
  \llap{$^d$} Argonne Leadership Computing Facility, 9700 S. Cass
  Avenue, Argonne, IL 60439, USA}

      \abstract{We present a new multigrid solver that is suitable
        for the Dirac operator in the presence of disordered gauge
        fields. The key behind the success of the algorithm is an
        adaptive projection onto the coarse grids that preserves the
        near null space. The resulting algorithm has weak dependence
        on the gauge coupling and exhibits mild critical slowing down in
        the chiral limit. Results are presented for the Wilson Dirac
        operator of the 2d U(1) Schwinger model.}

\FullConference{The XXV International Symposium on Lattice Field Theory\\
		 July 30-4 August 2007\\
		 Regensburg, Germany}

\begin{document}

\section{Introduction}
\secspace
The most demanding computational task in lattice QCD simulations
consists of the calculation of quark propagators, which are needed
both for generating gauge field configurations with the appropriate
measure and for the evaluation of most observables.  This calculation
consists of solving a very large system of linear equations,
\begin{equation}
D(U) \psi = \chi,
\label{eq:Dpsichi}
\end{equation}
where $\psi$ is the quark propagator, $\chi$ is the source term and
$D(U)$ is the discretized the Dirac operator matrix, with elements
dependent on the gauge field background $U$.

In the language of applied mathematics, Eq.~(\ref{eq:Dpsichi}) is a
discretized elliptic partial differential equation (PDE).  For
definiteness,
\begin{displaymath}
  D_{x,y} = - \frac{1}{2} \displaystyle \sum_{\mu=1}^{d} \bigl( (1-\gamma_\mu)U_x^\mu\, 
  \delta_{x+\hat\mu,y}\, + (1+\gamma_\mu)U_{x-\hat\mu}^{\mu \dagger}\, \delta_{x-\hat\mu,y}\bigr) + 
  (2d + m)\delta_{x,y}
\end{displaymath}
is the discretized Dirac operator describing a fermion in \(d\)
dimensions with mass \(m\) in the Wilson discretization of the Dirac
equation.  In the full 4 dimensional QCD problem (in volume \(V\)) the
matrices \(\gamma_\mu\) are the \(4\times4\) Dirac spin matrices and
\(U\) is the \(SU(3)\) gauge field.  It is this formulation that we
concentrate upon, however, we point out that many of the problems
encountered in solving this equation extend to other
formulations.

For any realistic QCD calculation the size of the matrix in
Eq.~(\ref{eq:Dpsichi}) is too large for a direct solver and iterative
Krylov-space methods must be used.  As the quark mass is brought
towards zero, the condition number of the matrix diverges and hence so
does the number of iterations until the desired convergence.  This
scaling with the mass is commonly referred to as critical slowing
down.

It has been known for some time that the multigrid (\mg) approach is
optimal when solving systems of the form \(Ax=b\), where \(A\) is the
sparse matrix that arises from the discretization of continuum
differential equations, \(b\) is a source vector and \(x\) is the
desired solution vector.  Here discretizations on successively coarser
(blocked) grids are used to accelerate the solver and this approach is
known to remove critical slowing down~\cite{Brandt:1977}.

One exception to the above statement is in solving the Dirac operator
in lattice QCD: here the nature of the underlying gauge field in the
Dirac operator has proven to be especially resistant to \mg.  Previous
attempts at \mg\ solvers have relied on renormalization group
arguments to define the coarse grids without realizing why the \mg\
approach succeeds, and this has invariably led to failure as the
physically interesting regime is approached
~\cite{Brower:1991xv,Lauwers:1992cp}~\footnote{We note, however, that
  recent progress has been made in the use of renormalization group to
  define a coarse Dirac operator, which may render this statement
  erroneous~\cite{Borici:2007ft}.}.  In this work we demonstrate an
\mg\ algorithm for the Dirac operator normal equations, i.e., the
positive definite operator given by
\[
A= D^\dagger D,
\]
that is shown to work in all regimes and vastly reduces the notorious
critical slowing of the solver as the renormalized fermion mass is
brought to zero.  We do so in the context of a 2-dimensional system
with $U(1)$ gauge field (Schwinger model).  This system captures many
of the physical properties (confinement, chiral symmetry breaking,
existence of non-trivial topological sectors) of the more complex
4-dimensional QCD.

\secspace
\section{Multigrid}
\secspace
The original formulation of \mg\ is best viewed with the example of
the free Dirac operator.  Multigrid solvers are based on the
observation that stationary iterative solvers (e.g., Jacobi,
Gauss-Seidel) are only effective at reducing local error components
leaving slow to converge, low wave-number components in the error.
For the free Dirac operator these slow modes will be geometrically
smooth and can be accurately represented on a coarser grid using
simple linear averaging.  However, on the coarse grid these low
wave-number error components become modes of shorter range and so
relaxation should be effective at removing them.  This process can
continue, moving to coarser and coarser grids until the degrees of
freedom have been thinned enough to solve the system exactly.  The
solution is then promoted back to the finest grid using linear
interpolation, where at each level relaxation is applied to the
correction vector to remove any high wave-number error components that
were introduced.  This process is known as a V-cycle~\cite{Brandt:1977}
and can be used as a solver in its own right, or more effectively as a
preconditioner for a Krylov method e.g., conjugate gradients (CG).

Before continuing we introduce the notation where the degree of
coarseness is represented by the integer \(l\), where \(l=1\)
represents the finest grid (i.e., where our actual problem is defined)
and \(l=L\) is the coarsest grid in an \(L\)-level \mg\ algorithm.
The operator used to promote a coarse grid vector on grid \(l+1\) to
the adjacent fine grid \(l\) is known as the prolongator
\(P^{(l,l+1)}\), and the converse operator is the restriction operator
\(Q^{(l+1,l)}\) which projects a fine grid vector onto the adjacent
coarse grid.  Typically the Galerkin definition is used to define the
coarse grid operator~\cite{Brandt:1977},
\begin{equation}
  A^{(l+1)} = Q^{(l+1,l)} A^{(l)} P^{(l,l+1)} = P^{(l,l+1)\dagger} A^{(l)} P^{(l,l+1)},
\label{eq:Galerkin}
\end{equation}
where we have defined the restriction operator as \(Q=P^\dagger\).
This guarantees the coarse grid operator is Hermitian positive
definite.  That the Galerkin definition is the optimum definition for
\(A\) can be found by minimizing the error of the coarse grid
corrected solution vector in the \(A\)-norm.  Apart from the coarsest
level which is an exact solve, each level of the V-cycle is the
following
\begin{enumerate}
\item Relax on the input vector, \(x^{(l)} = R^{(l)\dagger} b^{(l)}\),
  where \(R^{(l)\dagger}\) is a suitable relaxation
  operator.\footnote{The relaxation operator need not be Hermitian for
    the entire V-cycle to be Hermitian: the post-relaxation operator
    need only be the Hermitian conjugate to pre-relaxation.}
\item Restrict the resulting residual to the next coarsest grid, 
  \(r^{(l+1)} = P^{(l,l+1)\dagger} (b^{(l)}-A^{(l)}x^{(l)})\).
\item Apply the \(L=l+1\) V-cycle on the coarse residual, 
  \(e^{(l+1)} = V^{(l+1)} r^{(l+1)}\).
\item Correct the current solution with coarse grid correction,
  \(x^{(l)} = x^{(l)} + P^{(l,l+1)} e^{(l+1)}\).
\item Relax on the final residual, \( x^{(l)} = R^{(l)} (b^{(l)} - A^{(l)}x^{(l)})\).
\end{enumerate}
Written explicitly in terms of operators the \(l^{th}\) level of the
V-cycle thus takes the following form
\begin{equation}
  V^{(l)} = R^{(l)} + R^{(l)\dagger} + R^{(l)}A^{(l)}R^{(l)\dagger} +
  \Big[ (1 - R^{(l)}A^{(l)}) P^{(l,l+1)} V^{(l+1)}
  P^{(l,l+1)\dagger} (1 - A^{(l)} R^{(l)\dagger}) \Big].
\label{eq:vcycle}
\end{equation}
In this form the Hermiticity of the V-cycle is obvious (a necessary
condition if we are to use the V-cycle as a CG preconditioner).  The
cost of applying the \mg\ V-cycle becomes apparent from this explicit
form: on each level we must apply the operator \(A^{(l)}\) a total of
\(2\nu + 2\) times for each \(l\), where \(\nu\) is the number of
steps within the relaxation operator.

The problem in the application of the above procedure in the presence
of a non-trivial gauge field is that the eigenvectors responsible for
slow convergence are no longer low wave-number modes with
geometrically smooth variation.  They are instead modes that exhibit
localized lumps, typically extending over several lattice spacings.
An approach that was followed in \cite{Brower:1991xv} was to impose
Dirichlet boundary conditions along the block boundaries, and to use
the low modes of resulting blocked operator to define the prolongator.
This approach is bound to produce only a limited advantage since the
lumps of the low modes can span between several such blocks, so the
blocked operator will not possess this property.  Indeed in
\cite{Brower:1991xv} some acceleration was obtained but critical
slowing down was found to return after the correlation length of the
pion \(\mu^{-1}\) exceeded the correlation length \(l_\sigma\) of the
underlying gauge field.

\secspace
\section{Adaptive Multigrid}
\label{sec:adaptive}
\secspace
A breakthrough in the application of multiscale methods to more
complex problems occurred with the discovery of adaptive \mg\
techniques~\cite{Brezina:2004, Brannick:2006}.  In the adaptive
algorithm the \mg\ process itself defines the appropriate prolongator
by an iterative procedure which we now concisely describe.

In the first pass, one uses relaxation alone to solve the homogenous
problem \(Ae=0\) with a randomly chosen initial error vector.  After a
certain number, \(\nu\), of relaxation steps, the relaxation procedure, which we
symbolically represent by
\begin{equation}
e \to e^\prime=(I-\omega A)^\nu e \equiv (I- \omega D^\dagger D)^\nu e,
\label{eq:relax}
\end{equation}
produces an \(e^\prime\) that essentially belongs to the space spanned
by the slow modes, so $e^\prime$ is now used to define a first
approximation to the prolongator $P$.  One blocks the variables of the
original lattice into subsets, which we denote by $S_j$.  From
$e^\prime$ we construct the vectors $e^\prime_j$, which are identical
to $e^\prime$ within $S_j$ and $0$ outside $S_j$, and the vectors of
unit norm $v_{1j}=e'_j /\vert e'_j \vert$.  The extra ``1'' index in
$v_{1j}$ has been introduced for a discussion that follows.  The
prolongator $P^{(1,2)} \equiv P^{(1,2)}_{i,j}$ which maps a vector
$\psi_j^{(2)}$ in the coarse lattice, indexed by $j$, to the original
lattice, where $i$ denotes collectively the site, spin and possible
internal symmetry indices, is then defined by
\begin{equation}
P^{(1,2)}_{i,j}=v_{1j,i},
\label{eq:P1}
\end{equation}
where we have made explicit the fine lattice indices of $v_{1j}$.

There are variations on how to block the fine lattice, i.e.,~how to
define the sets $S_j$.  In the so called ``algebraic adaptive MG'' one
partitions the fine lattice into subsets on the basis of the magnitude
of the matrix elements of $A$.  Since such matrix elements in lattice
gauge theories are typically of uniform magnitude, differing rather in
phase or, in a broader sense, in orientation within the space of gauge
transformations, we chose instead to partition the lattice
geometrically into fixed blocks of neighboring lattice sites,
specifically $4\times 4$ squares in our study of the Schwinger model.
Maintaining a regular lattice on coarse levels will allow more
efficient parallel code with exact load balancing.

Another refinement of the technique consists of applying a simple
Richardson iteration to the vectors $v_{1j}$ before defining the
prolongator.  The choice of damping parameter in this smoothing
procedure is chosen to minimize the condition number of the resulting
coarse grid operator.  The term ``smoothed aggregation'' is used for
this.  Thus our overall technique can be referred to as ``geometric
adaptive smoothly aggregated MG''.

We come now to the crux of the adaptive \mg\ method.  We use the
prolongator defined above (Eq.~(\ref{eq:P1})) to implement a standard
\mg\ V-cycle and apply it, like relaxation before, to a randomly
chosen error vector.  There are two possibilities.  Either the V-cycle
reduces the error with no sign of critical slowing down or some large
error, $e^{\prime \prime}$, survives the cycle.  In the first case, of
course, one need not proceed: the \mg\ procedure works as is.  In the
second case, we define another set of vectors $v_{2j}$ over the coarse
lattice by restricting $e^{\prime \prime}$ to the subsets $S_j$,
making the new vectors orthogonal to the vectors $v_{1j}$ and
normalizing them to 1.  The smoothed aggregation procedure is now
applied to the set $v_{sj}\equiv (v_{1j}, v_{2j})$.  A new prolongator
is defined by projecting over these vectors
\begin{displaymath}
P_{i,sj}^{(1,2)}=v_{sj,i},
\end{displaymath}
where the index $s$ (taking values $1,2$) can be considered
as an intrinsic index over the coarse lattice.

The procedure described in the above paragraph is repeated as
necessary, until the repeated application of a V-cycle reduces a
random initial error sufficiently without critical slowing down.  The
method works if critical slowing down is eliminated with a few
iterations of the adaptive procedure.  If this occurs with $M$ vector
sets, then the coarse lattice will carry $M$ degrees of freedom per
site.  As with all \mg\ methods, the procedure is recursive and it can
be used to define further coarsenings.

\secspace
\section{Results}
\secspace
In testing this algorithm for lattice QCD we generated quenched $U(1)$
gauge field configurations on a $128 \times 128$ lattice with the
standard Wilson gauge field action
\begin{displaymath}
S=\sum_{x,\nu<\mu} \beta\, {\rm Re}\, U_x^{\mu \nu} \equiv
\sum_{x,\nu<\mu} \beta\, {\rm Re}\, U_x^{\mu} U_{x+\hat \mu}^{\nu}
U_{x+\hat \nu}^{\mu \dagger} U_x^{\nu \dagger}
\end{displaymath}
and periodic boundary conditions at $\beta=6$ and $\beta=10$.  These
two values of $\beta$ define correlation lengths for the gauge field
to be $l_\sigma = 3.30 $ and $l_\sigma = 4.35$ respectively, via the
area law for the Wilson loop: $ W \sim \exp[ -A/l^2_\sigma]$. For
comparison on these lattices, a fermion mass gap \(\hat{m} = m -
\rmsub{m}{crit} = 0.01\) corresponds to the pseudoscalar meson
correlation lengths \(\mu^{-1} = 6.4 \) and \(\mu^{-1} = 12.7 \)
respectively.~\footnote{All quantities are expressed in lattice
  units.}  In the 2-dimensional \(U(1)\) gauge theory, one can
identify a gauge invariant topological charge \(\hat{Q}\), which in the
continuum limit is proportional to the quantized magnetic flux flowing
through the system.  A gauge field with nonzero \(\hat{Q}\) corresponds to a
Dirac operator with exactly real eigenvalues and, hence, as the mass
gap is brought towards zero the condition number becomes infinite.
Thus, it is important to test both trivial (\(\hat{Q}=0\)) and non-trivial
(\(\hat{Q}\neq 0\)) gauge field topologies.

We blocked the lattice into $4 \times 4$ blocks and implemented the
adaptive \mg\ procedure described above.  We used a degree 2
polynomial smoother for our relaxation procedure, where the
coefficients were chosen by running two iterations of an underrelaxed
minimum residual solver (\(\omega=0.8\)) and subsequently held fixed
(hence, for our choice of smoother \(R=R^\dagger\)).  The coarsening
procedure was repeated twice maintaining $M=8$ vectors in all
coarsenings, down to an $8 \times 8$ lattice, over which the equations
were solved exactly.  For each gauge field we performed the set up
procedure for the \mg\ preconditioner for the lightest mass parameter
only, and reused the vectors for each heavier mass. We used this
constructed V-cycle as a preconditioner for CG where the operator
defined in Eq.~(\ref{eq:vcycle}) is applied at each iteration to the CG
direction vector (here on referred to as MG-CG).

\begin{figure}
  \hfill
  \begin{minipage}[t]{0.47\textwidth}
    \includegraphics[width=1.05\textwidth]{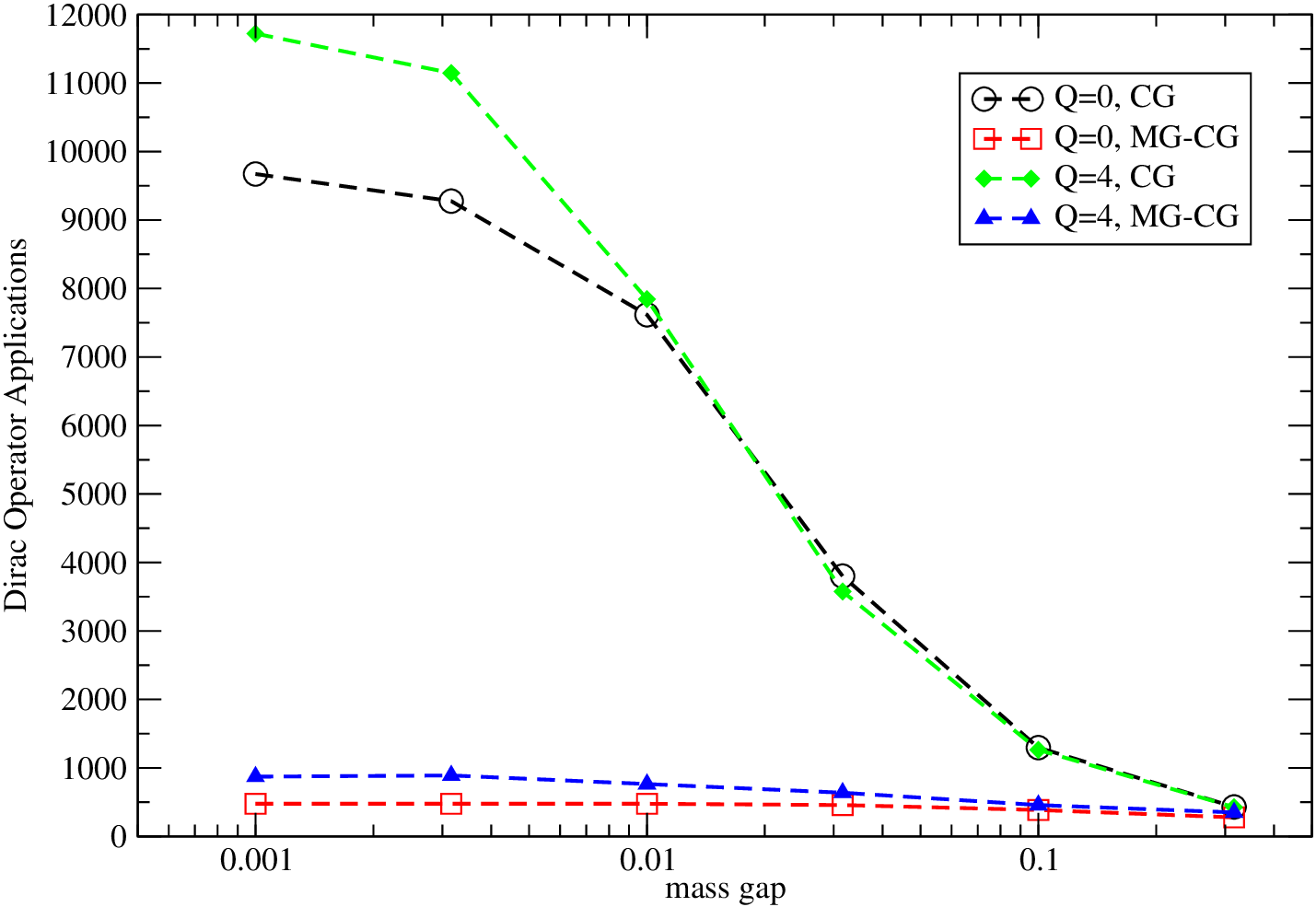} 
    \caption{Number of Dirac operator applications of CG vs.~MG-CG as
      function of the mass gap at $\beta =6$ (point source, relative
      residual \(|r|=10^{-14}\)).}
    \label{fig:beta6} 
  \end{minipage}
  \hfill
  \begin{minipage}[t]{0.47\textwidth}
    \includegraphics[width=1.05\textwidth]{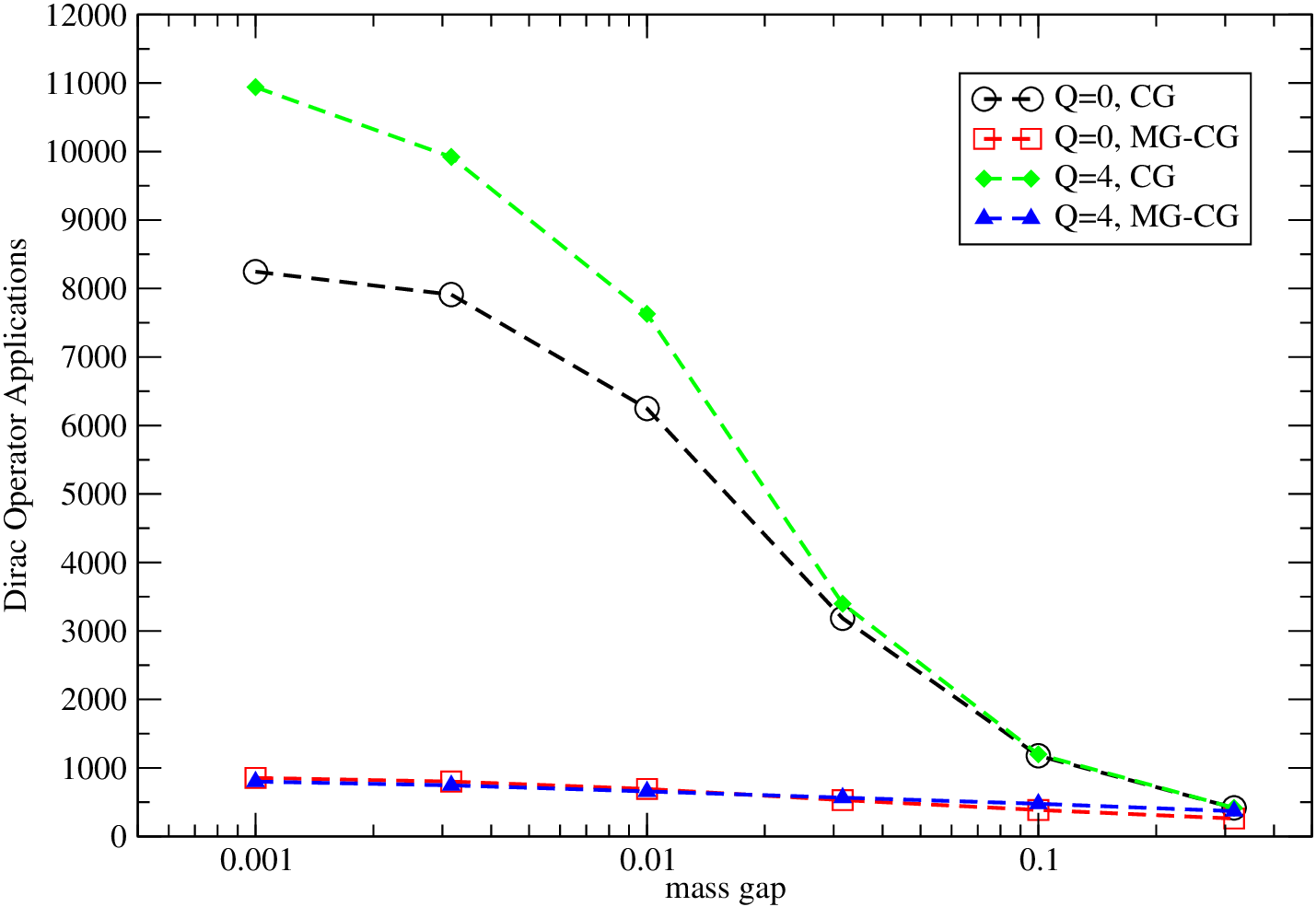} 
    \caption{Number of Dirac operator applications of CG vs.~MG-CG as
      function of the mass gap at $\beta =10$ (point source, relative
      residual \(|r|=10^{-14}\)).}
    \label{fig:beta10} 
  \end{minipage}
  \hfill
\end{figure} 

If one compares the number of CG iterations needed to achieve
convergence with or without \mg\ preconditioning, the gain obtained
with the \mg\ method is dramatic: for example, with \(\beta =6\),
\(\hat{m}=0.01\) and \(\hat{Q}=0\), it takes 3808 iterations to
achieve convergence with a straightforward application of the CG
technique, whereas it takes only 26 iterations using MG-CG.  However
this comparison does not take into account the fact that many more
operations per iterations must be performed when applying the \mg\
preconditioner.  To achieve a more balanced comparison, in
Figs.~\ref{fig:beta6},~\ref{fig:beta10} we plot the total number of
applications of \(D\) and \(D^\dagger\) done on the fine lattice.
This reflects better the actual cost of the calculations (at each
iteration of MG-CG there are 6 applications of \(D^\dagger D\): 1
application in the outer CG, and 2 pre- and 2 post- coarsening
smoothing applications and 1 further application required to form the
residual).  We do not include the additional cost arising from the
coarse lattices since this is expected to be a small overhead, and has
not been optimized for our model calculation.  The advantage coming
from the use of the adaptive \mg\ technique is still very dramatic:
critical slowing down, if not totally eliminated, is very
substantially reduced and there is no slow down as as the pion
correlation length exceeds the gauge field correlation length.  These
results are for point sources, however, we tried a variety of
different source vectors for this analysis (e.g., Gaussian noise,
\(Z_4\) noise) and found very little dependence of MG-CG performance
on the source vector.

From the point of view of computational complexity, one should also
take into account the cost of setting up the \mg\ preconditioner,
i.e.,~of constructing the prolongator $P$.  This cost is however
heavily amortized, to the point of being negligible, if, as is often
the case, one must apply the solver to systems with multiple given
vectors (for example, solving for all color and spin components of a
quark propagator or, in the calculation of disconnected diagrams where,
\(O(1000)\) inverses are required to estimate the trace of the inverse
Dirac operator).

\secspace
\section{Conclusion}
\secspace
Our results, albeit for now limited to a 2-dimensional example,
provide a clear indication that adaptive \mg\ can be made to work with
the lattice Dirac operator.  What appears to be at the root of its
success is that, although the modes responsible for slow convergence
of the Dirac solver on a fine lattice are not low wavenumber
excitations, like in the free case, their span can be well
approximated by a set of vectors of limited dimensionality on the
blocks that define the coarse lattice.\footnote{The observation that
  the space of slow modes may be of limited span is also at the root
  of a method recently proposed by L\"uscher in
  Ref.~\cite{Luscher:2007se}, although the technique there is quite
  different from the one we follow.}  Earlier
attempts~\cite{Brower:1991xv,Lauwers:1992cp} failed to eliminate
critical slowing down when the pseudoscalar length exceeded the
disorder length of the gauge field: $\mu^{-1} > l_\sigma$.  Adaptive
\mg\ finds the coarse subspaces through the iterative application of
the method itself.  It is of course crucial that the approximation to
the space of slow modes can be achieved with a small number of vectors
on the individual blocks, otherwise the application of the method
would not be cost effective.  But this appears to be the case in the
examples we studied and, if the results hold true in general, adaptive
\mg\ has the potential of substantially speeding up lattice QCD
simulations as the increase of available computational power leads one
to consider ever larger lattices.

Current research is focussed on applying this adaptive \mg\ algorithm
to the Dirac operator directly, as opposed to the normal equations.
Here we are motivated to do so because of the reduced condition number
and the increased sparsity of the operator.  There are added
complications because the Dirac operator isn't Hermitian (this
requires that the restriction and prolongation operators are defined
using left and right vectors respectively), however, initial progress
is extremely promising.  The application of the method to large
4-dimensional systems is in progress,

Acknowledgments.  This research was supported in part under NSF grant
PHY-0427646 and DOE grants DE-FG02-91ER40676 and DE-FC02-06ER41440.


\begin{thebibliography}{99}




\bibitem{Brandt:1977}
A.~Brandt.
\newblock {\em Math. Comp.}, 31:333--390, 1977.

\bibitem{Brower:1991xv}
Richard~C. Brower, Robert~G. Edwards, Claudio Rebbi, and Ettore Vicari.
\newblock {\em Nucl. Phys.}, B366:689--705, 1991.

\bibitem{Lauwers:1992cp}
P.~G. Lauwers and T.~Wittlich.
\newblock {\em Int. J. Mod. Phys.}, C4:609--620, 1993.

\bibitem{Borici:2007ft}
  A.~Borici,
  2007 (arXiv:0704.2341 [hep-lat]).

\bibitem{Brezina:2004}
M.~Brezina, R.~Falgout, S.~MacLachlan, T.~Manteuffek, S.~McCormick, and
  J.~Ruge.
\newblock {\em Siam J. Sci. Comput.}, 25:1896--1920, 2004.

\bibitem{Brannick:2006} 
  J.~Brannick, M.~Brezina, D.~Keyes, O.~Livne, I.~Livshits,
  S.~MacLachlan, T.~Manteuffel, S.~McCormick, J.~Ruge, and L.~Zikatanov.
  \newblock{\em Lecture Notes in Computational
    Science and Engineering}, 55:499--506, 2006.

\bibitem{Luscher:2007se}
M.~L\"{u}scher.
\newblock 2007 (arXiv:0706.2298 [hep-lat]).

\end{thebibliography}
\end{document}